\documentclass[preprint,aps,tightenlines,showpacs,superscriptaddress]{revtex4}
\usepackage[dvips]{graphicx}
\usepackage{dcolumn}
\usepackage{bm}
\usepackage{epsfig}
\usepackage{color}

\begin{document}

\title{Neutrino deuteron reaction in the heating mechanism of core-collapse supernovae}

\vspace{0.5cm}


\author{S. X. Nakamura}
\affiliation{Instituto de F\'isica, Universidade de S\~ao Paulo,
C.P. 66318, 05315-970, S\~ao Paulo, SP, Brazil}

\author{K. Sumiyoshi}
\affiliation{Numazu College of Technology, Ooka 3600, Numazu,
Shizuoka 410-8501, Japan}

\author{T. Sato}
\affiliation{Department of Physics, Osaka University, Toyonaka,
Osaka 560-0043, Japan}

\begin{abstract}
We examine a potential role of the neutrino deuteron reactions 
in the mechanism of supernova explosion by evaluating 
the energy transfer cross section for the neutrino heating.  
We calculate the energy loss rate due to the neutrino absorptions 
through the charged-current process as well as the neutrino scattering 
through the neutral-current process.  
In so doing, we adopt a detailed evaluation of cross sections 
for the neutrino deuteron reactions with the phenomenological 
Lagrangian approach.  
We find the energy transfer cross section for the deuteron
is larger than those for $^{3}$H, $^{3}$He and $^{4}$He for 
neutrino temperatures (T$_\nu$ $\sim 4$ MeV) relevant to supernova core.  
Because of the low energy threshold for the deuteron breakup, 
the energy transfer rate rapidly increases from low temperature, 
T$_\nu$ $\sim 1$ MeV.  
This suggests that the neutrino deuteron reactions may 
contribute effectively to the heating mechanism 
during the dissociation of irons into light elements and 
nucleons in the shocked material of supernova core.  
\end{abstract}

\pacs{21.45.Bc, 23.40.Bw, 25.30.Pt, 26.50.+x, 97.60.Bw}

\maketitle

\section{introduction}

The reactions between neutrinos and nuclei play important roles
in various phenomena in physics and astrophysics.
The neutrino-nucleus interactions are crucial in physics of
core-collapse supernovae through neutrino scattering, absorption and emission.
Reliable evaluation of the rates of neutrino-nucleus interactions
is necessary for the terrestrial neutrino detectors
such as Super-Kamiokande and SNO \cite{tkk,ying89,nsgk,netal,BCK,ando}, 
for predicting the amount of elements produced
in the nucleosynthesis 
through neutrino-processes \cite{haxton2,yoshida04,tsuzuki06}, 
and for understanding the mechanism of supernova explosion\cite{bethe85,haxton} 
after the core bounce of the gravitational collapse of massive stars.

In the neutrino heating mechanism, which is one of the key
issues for the successful supernova explosion,
the neutrino-nucleus interactions may contribute
to the energy deposition in addition
to the neutrino-nucleon interactions \cite{haxton,bruhax}.
In most of recent studies on supernovae, the shock wave launched
by the core bounce stalls on the way and needs extra
assistance to revive the outward propagation \cite{bur07,jan07}.
Although the absorption of neutrino on the nucleon is
the main mechanism of the energy deposition,
nucleus can be an extra agent of the energy deposition
through the absorption or scattering.
An advantage of nuclei as targets is that neutrinos interact
through the neutral-current (NC) as well as the charged-current (CC).
Since the neutrinos of $\mu$- and $\tau$-types have relatively
higher average energies than those of electron-type,
they can have larger cross sections than those from
the CC reactions for the low-energy electron-type neutrinos.

The neutrino heating through the neutrino-nucleus interactions
has been proposed by Haxton and the average energy
transfer cross section has been evaluated
for the representative case of $^{4}$He and $^{56}$Fe \cite{haxton}.
Since the matter from outer layer free-falls down to the
shock wave and is dissociated into $^{4}$He and nucleons
due to the shock heating, those two representative species
are potential targets of neutrinos streaming from the central core.
Although its influence on the supernova dynamics has been studied
by numerical simulations under the spherical symmetry
with those reaction rates \cite{bruhax}, it was
reported that their effect is too small to affect
significantly the dynamics.
In a recent study on the standing accretion shock-wave
instability (SASI), Ohnishi et al. \cite{ohnishi} have explored 
how the inelastic scattering of neutrinos
with $^{4}$He influence
the outcome of shock dynamics
by performing numerical simulations in two dimensions.
An effect on the growth of instability toward
the explosion is found only for the critical condition
in which the dynamics of shock wave is marginal to the explosion
and exploratory cases with enhanced reaction rates.
The effect depends crucially on the cross section
of the neutrino-$^{4}$He scattering and detailed evaluation
based on theoretical models have been done recently \cite{tsuzuki06,gb}.  

Recently, several authors pointed out 
the appearance of light elements (A $\le$ 3)
other than $^{4}$He in supernova environment,
and its potential role on the supernovae
through the neutrino interactions\cite{cghsb,sr,arcones}.
O'Connor et al. \cite{cghsb} studied the abundance of
tritons and $^{3}$He in dense matter and evaluated
the neutrino reaction rates.
The A=3 nuclei can be more abundant than $^{4}$He
and nucleons, and its energy transfer rate can be
larger than that of $^{4}$He for certain conditions
of proton fraction, density and temperature.
Arcones et al. \cite{arcones} used a statistical model and 
studied the abundance of light elements in the neutrino-driven flows
after the explosion.
They demonstrated that the neutrino reactions
on the light elements can have an influence on the spectra
of neutrinos emitted from the surface of proto-neutron stars.

So far, attention
has been paid to the neutrino reactions on A=3, 4 nuclei only.
Sumiyoshi and R{\"o}pke \cite{sr} studied
the composition of light elements in supernovae by
including the deuteron, A=3, 4 nuclei and others
in many-body calculations (See also \cite{arcones}),
and found that the deuteron
can also appear abundantly in some regions of supernova
environment.
Since the cross section of the neutrino-deuteron reaction
is much larger than those for $^{3}$H, $^{3}$He and $^{4}$He,
the energy deposition through the deuteron may significantly contribute
to the neutrino heating during the dissociation of iron 
nuclei into light elements and nucleons.

In this paper, we examine the possible role of
the deuteron by evaluating the neutrino heating rates
through the neutrino-deuteron ($\nu d$) interactions.
For the estimation of the $\nu d$ reaction cross sections,
we adopt the standard nuclear physics approach(SNPA)\cite{nsgk,netal}.
In SNPA, we consider the nuclear current consisting of
one-body impulse terms and two-body exchange-current terms,
and evaluate their matrix element with the nuclear wave
functions generated with a high-precision NN potential.
The reliability of our SNPA has been extensively tested by comparisons of
calculated observables with available data of photo-processes, and also
from a good agreement with the results of the effective field theory 
approach\cite{netal,ando}
for the solar neutrino energy.
We calculate the average energy transfer rates (and the heating rate)
by using the Fermi distribution of the neutrino energy,
and examine an  effect
 of the $\nu d$ reaction on the supernova dynamics
in the heating region.
We have in our mind here the typical situation realized in the region
between the stalled shock wave and the proto-neutron
star surface, where neutrinos are emitted with
spectra close to the one under the equilibrium
parametrized by the local temperature.
We consider the neutrino reactions on the deuteron
through both the CC and NC,
covering the deuteron breakup and scattering.

We show that the energy transfer rates through
the $\nu d$ reactions are much larger than
those  of $^{3}$H, $^{3}$He and $^{4}$He
for the typical range of the neutrino temperature,
especially at low temperature (energy)
because of the lower breakup threshold.
This suggests that the deuteron can contribute,
as an extra agent, to the neutrino heating
in supernova dynamics.
The current finding urges ones to examine its effects on the supernova explosion
by implementing the $\nu d$ reactions 
as well as a detailed 
treatment of the mixture of light elements 
in supernova simulations.

\section{Formulation}

We are concerned with the  charged-current (CC)
and neutral-current (NC) $\nu(\bar{\nu})d$ reactions listed
in the following equations.
\begin{eqnarray}
\nu_e + d \rightarrow e^- + p + p \ \ [\nu CC] \\
\bar{\nu}_e + d \rightarrow e^+ + n + n \ \ [\bar{\nu} CC] \\
\nu + d \rightarrow \nu + p + n \ \ [\nu NC] \\
\bar{\nu} + d \rightarrow \bar{\nu} + p + n \ \ [\bar{\nu} NC] \\
\nu/\bar{\nu} + d \rightarrow \nu/\bar{\nu} + d \ \ [\nu/\bar{\nu}scatt]
\end{eqnarray}
In the supernova environment, 
the CC reactions act as an absorber of the neutrino,
and the neutrino energy is deposited to the rest of compositions 
of supernova matter.  
For the NC reactions, the energy difference between the
incoming and outgoing neutrinos is the energy transfer 
to the matter.  

The interaction Hamiltonian 
for semileptonic weak processes is
given by the product of 
the hadron current ($J_{\lambda}$) and 
the lepton current ($L^{\lambda}$) as
\begin{eqnarray}
H_W^{CC} & = & \frac{G_F' V_{ud}}
{\sqrt{2}}\int d\bm{x}  [
   J_{\lambda}^{CC}(\bm{x})L^{CC,\lambda}(\bm{x}) +
    \mbox{h. c.}] \ , \label{eq_Ham-CC}
\end{eqnarray}
for the CC process and
\begin{eqnarray}
H_W^{NC} & = & \frac{G_F'}{\sqrt{2}}\int d\bm{x}
\  [J_{\lambda}^{NC}(\bm{x})L^{NC,\lambda}
    (\bm{x})+\mbox{h. c.}] \ , \label{eq_Ham-NC}
\end{eqnarray}
for the NC process.
Here $G_F'$ is the Fermi coupling constant,
and $V_{ud}$ is CKM matrix element.
For the weak coupling constant we adopt
$G_F'=1.1803 \times 10^{-5}\mbox{GeV}^{-2}$ \cite{netal}.  
The CKM matrix element is taken to be $V_{ud} = 0.9740$.

The hadronic CC is written as
\begin{eqnarray}
J_{\lambda}^{CC}(\bm{x}) & = & 
V_{\lambda}^{1 \pm i 2}(\bm{x}) +
 A_{\lambda}^{1 \pm i 2}(\bm{x}) ,
\end{eqnarray}
where $V_{\lambda}$ and $A_{\lambda}$ 
denote the vector and axial-vector currents, respectively.
The  $J_\mu^{1 \pm i2} $ for the $\nu/\bar{\nu}$-reaction denotes
$J_\mu^1 \pm i J_\mu^2$,
where $J^i$ is the $i$th component of the isovector current.
The hadronic NC is given by
\begin{eqnarray}
J_{\lambda}^{NC}(\bm{x}) 
&=& (1-2 \sin^2 \theta_W )V_{\lambda}^{3}(\bm{x}) +
A_{\lambda}^{3}(\bm{x}) -2 \sin^2 \theta_W 
V_{\lambda}^{s}(\bm{x}) , \label{eq_NC-current}
\end{eqnarray}
where $\theta_W$ is the Weinberg angle and
$V_{\lambda}^{s}$ is the isoscalar part 
of the vector current.
The lepton currents, $L^{CC,\lambda}$ and $L^{NC,\lambda}$,
are well known.

The nuclear current consists of 
one-nucleon impulse approximation (IA) terms
and two-body meson exchange-current (EXC) terms.
The IA current is determined by the single-nucleon
matrix element of $J_\lambda$ with the standard parametrization.
For the axial vector EXC ($A_{EXC}$), we consider the pion-pair current,
rho-pair current, pion- and rho-exchange $\Delta$ currents, $\pi-\rho$ current
following Refs. \cite{crsw,schi}.  The strength of $A_{EXC}$ is adjusted
to reproduce the experimental value
of the triton beta decay rate. Regarding the vector EXC, we take into account
the pion-pair, pionic and pion- and rho-exchange $\Delta$ currents.
As discussed in \cite{nsgk}, 
the model of the nuclear vector current
leads to $n p \rightarrow d \gamma$ total cross sections that agree 
very well with the experimental values. 
The expressions of IA, EXC and the coupling constant 
used in this work are given in Ref.~\cite{nsgk,netal}. 

The cross sections for the
 $\nu/\bar{\nu}(k) + d(P) \rightarrow l(k') + N_1(p_1') + N_2(p_2')$ 
in the laboratory
system are calculated following the standard procedure.
We obtain the cross section for the CC reaction as
\begin{eqnarray}
d\sigma = \sum_{\bar{i},f} 
          \frac{\delta^{(4)}(k+P-k'-P')}{(2\pi)^5} 
          \frac{G_F^{\prime 2} V_{ud} }{2}\,
           F(Z,E'_{\ell})\,\,|l^ \lambda j_{\lambda}^{CC}|^2 
          d\bm{k'} d\bm{p'}_1d\bm{p'}_2, 
   \label{eq_cs-CC}
\end{eqnarray}
where we have included the Fermi function $F(Z,E_l')$
to take into account the Coulomb interaction between the electron
and the nucleons in the final state.
The cross section for the NC reaction is written as
\begin{eqnarray}
d\sigma = \sum_{\bar{i},f} 
          \frac{\delta^{(4)}(k+P-k'-P')}
     {(2\pi)^5} \frac{G_F^{\prime 2}}{2}|l^
          \lambda j_{\lambda}^{NC}|^2 d\bm{k'} 
           d\bm{p'}_1d\bm{p'}_2.\label{eq_cs-NC}
\end{eqnarray}
We have used the matrix elements $l^\lambda$ and $j^\lambda$ defined as
\begin{eqnarray}
j_\lambda & = & <NN(P')| J_\lambda(0)|d(P)>, \\
l_\lambda & = & <l(k') | L_\lambda |\nu/\bar{\nu}(k)>.
\end{eqnarray}

The cross section for the neutrino-deuteron elastic scattering 
($\nu/\bar{\nu}(k) + d(P=0) \rightarrow \nu/\bar{\nu}(k') + d(P')$)
is written as
\begin{eqnarray}
\frac{d\sigma}{dk'}
 & = & \frac{k'}{k}\frac{M_d}{\pi} G_F^{\prime 2} 4 \sin^4\theta_W
      [A \cos^2\theta_L/2 + B \sin^2\theta_L/2],
\end{eqnarray}
where $\theta_L$ is the scattering angle of the  neutrino in the laboratory
system and is determined from the energies of the neutrinos
in the initial($k$) and the final($k'$) states as
\begin{eqnarray}
\cos\theta_L = 1 - (k - k')M_d/(k k').
\end{eqnarray}
Since the deuteron is an iso-scalar object, only the iso-scalar vector
current of the hadronic NC contributes
to the $\nu d$ elastic scattering.
 Neglecting a small contribution from the strange form factor,
the matrix element of the hadronic current
can be expressed with the iso-scalar elastic electromagnetic form factors
of the deuteron.
The structure functions, $A$ and $B$, can be expressed
with the electromagnetic Coulomb monopole ($G_C$), magnetic dipole ($G_M$)
and quadrupole ($G_Q$) form factors of the deuteron as
\begin{eqnarray}
B & = & \frac{4}{3}\eta(1+\eta)G_M^2 \ , \\
A &=& G_C^2 + \frac{8}{9}\eta^2 G_Q^2 + \frac{2}{3}\eta G_M^2 \ ,
\end{eqnarray}
with $\eta = Q^2/4M_d^2$ and $Q^2=(\bm{k}-\bm{k}')^2-(E_\nu-E_\nu')^2$.
We use the IA current to calculate the iso-scalar form factors.
Since the dominant contribution to the elastic $\nu d$ scattering
is from $G_C$, effects of EXC is expected to be small
in the energy region of our interest.

In our numerical calculation,
we use the ANLV18 potential \cite{anlv18}  to generate the deuteron
and two-nucleon scattering wave functions. The NN partial waves up to
$J=6$ are included for the deuteron breakup reactions.

\section{Results and discussions}

\begin{figure}[h]
\includegraphics[width=8cm]{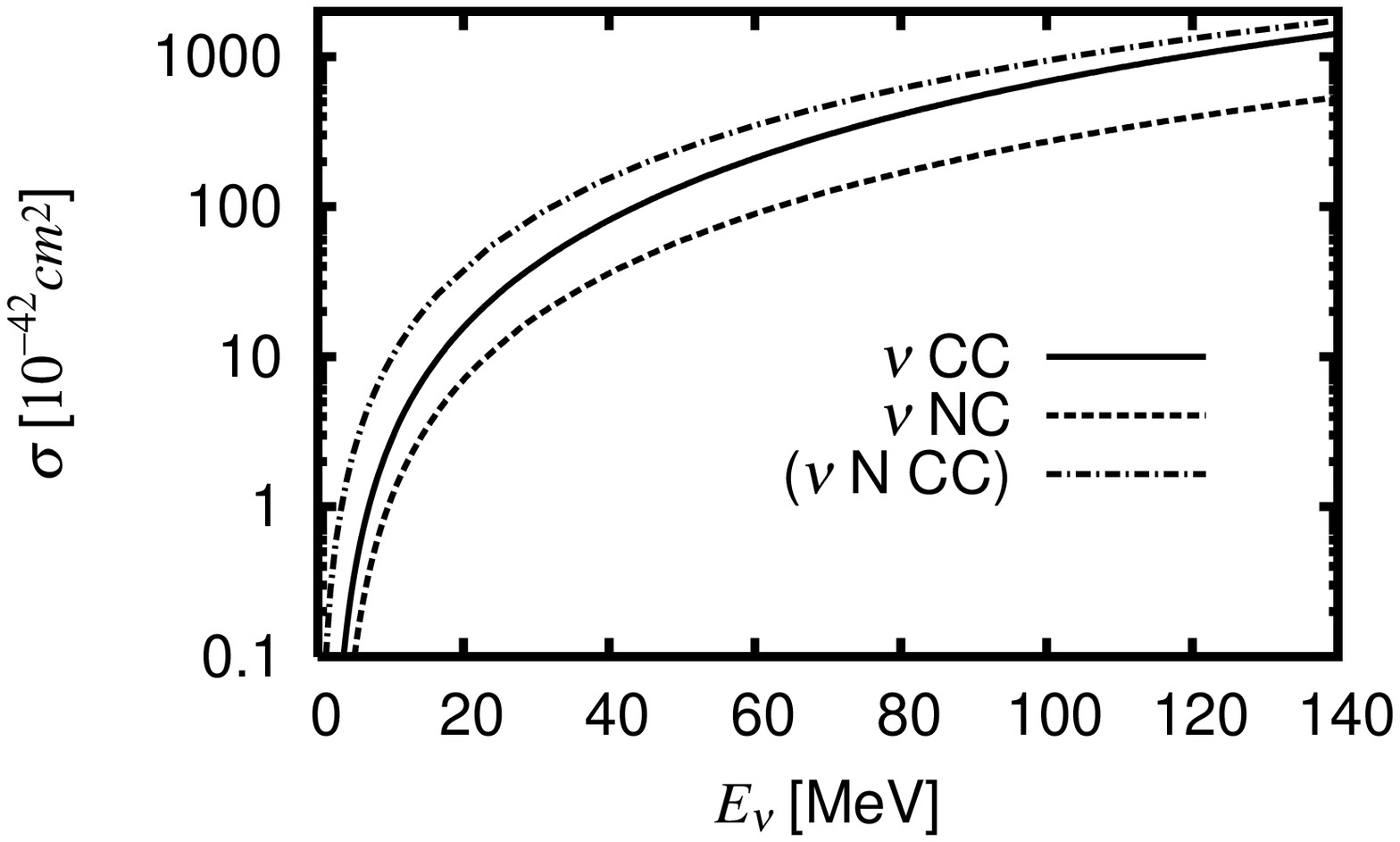}
\includegraphics[width=8cm]{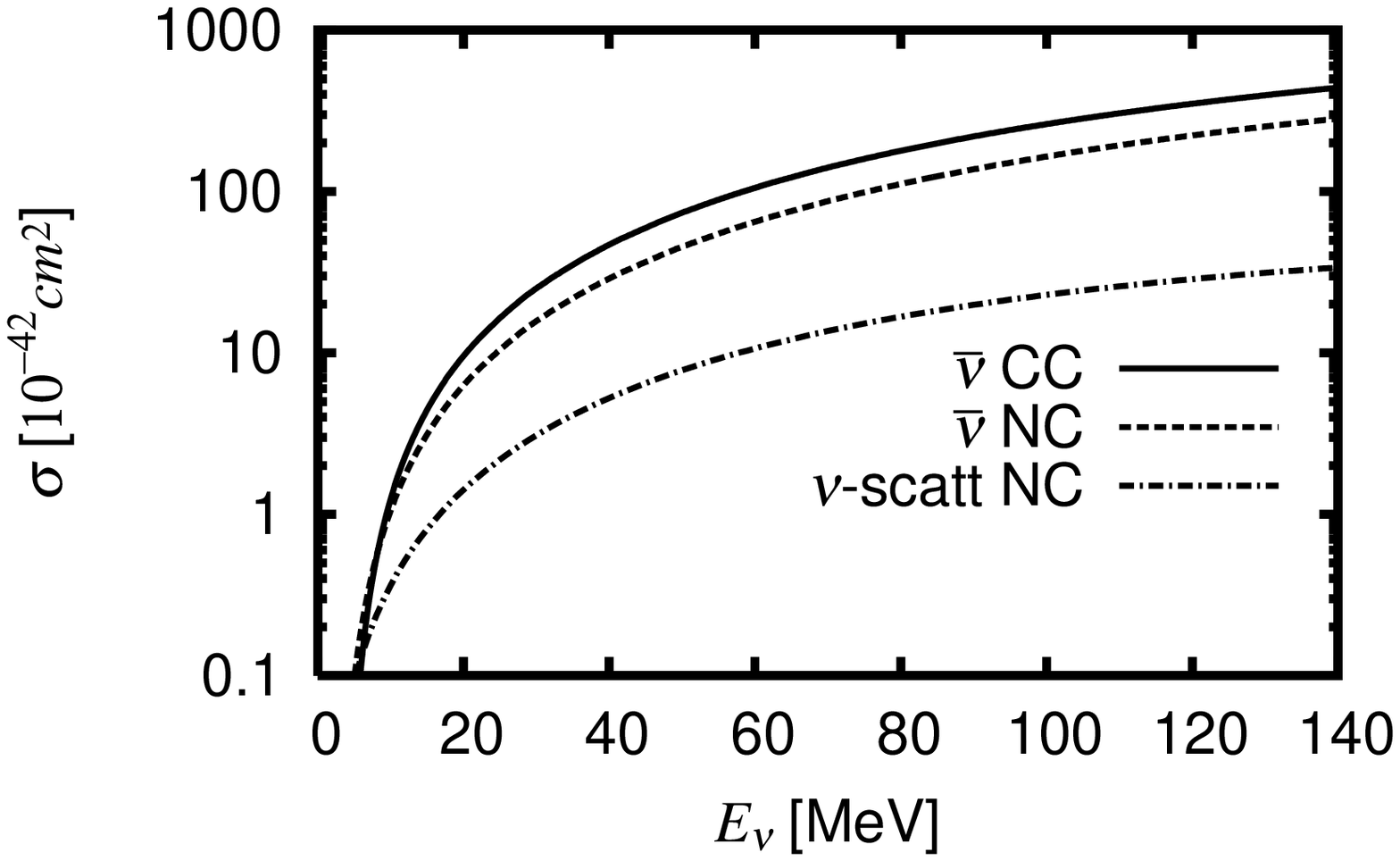}\\
\hspace*{2cm}(a)\hspace*{6cm}(b)
\caption{Total cross sections for the neutrino-deuteron reactions.
The solid and dotted curves show the 
cross sections for (a) $\nu_e d \rightarrow e^- p p$ ($\nu$CC) and 
$\nu d \rightarrow \nu p n$ ($\nu$NC), respectively and
(b) $\bar{\nu}_e d \rightarrow e^+ n n$ ($\bar{\nu}$CC) and 
$\bar{\nu} d \rightarrow \bar{\nu} p n$ ($\bar{\nu}$NC), respectively.
The dot-dashed curves show 
(a) the cross sections for 
$\nu + n \rightarrow e^- + p (\nu N CC)$ 
and (b) the cross sections for
$\nu + d \rightarrow \nu + d$ ($\nu-scatt$).}
\label{fig:cross}
\end{figure}

First we present the energy dependence of the total
cross sections for the reactions in Eqs.(1)-(5).
The total cross section for the neutrino
and anti-neutrino deuteron reactions are shown in Figs. \ref{fig:cross} (a)
and (b), respectively. 
The neutrino CC reaction ($\nu CC$) gives the largest
cross sections which are about $1/3$ ($1/2$) of the neutrino-nucleon
CC reaction at $E_\nu=10$($50$)MeV.
The $\nu d$ elastic cross section is very small compared with that of $\nu CC$.
In the high energy region
around the pion production threshold ($E_\nu \sim 140$MeV), 
the pion production cross sections can be safely neglected\cite{uno}.

An interesting quantity relevant to the supernova physics is
a thermal average of the energy transfer cross section defined by\cite{haxton}
\begin{eqnarray}
< \sigma  \omega >_{T_\nu}
 & = &  \int dE_\nu f(T_\nu,E_\nu) \sigma\omega(E_\nu) \ .
\end{eqnarray}
We assume here a Fermi-Dirac distribution for the neutrino,
having in mind the neutrino flux from the supernova core.  
At the temperature $T_\nu$ with zero chemical potential,
a neutrino with the energy $E_\nu$ distributes as
\begin{eqnarray}
f(T_\nu,E_\nu) & = & \frac{N}{T_\nu^3}\frac{E_\nu^2}{e^{E_\nu/T_\nu} +
 1} \ .
\end{eqnarray}
The energy transfer cross section $\sigma\omega(E_\nu)$
is evaluated by integrating 
the differential cross section multiplied by
the energy loss of the incident neutrino 
with respect to the energy of the final lepton:
\begin{eqnarray}
\sigma\omega(E_\nu) & = & \int dE_l' \frac{d\sigma}{dE_l'} E_\nu \ ,
\end{eqnarray}
for the CC reaction (absorption) and 
\begin{eqnarray}
\sigma\omega(E_\nu) & = & \int dE_\nu'
 \frac{d\sigma}{dE_\nu'}(E_\nu-E_\nu') \ ,
\end{eqnarray}
for the NC reaction (scattering).
This quantity is important for evaluating the neutrino 
heating behind the shock wave in the supernova.
The energy of the incident neutrino is transferred to the matter 
(deuterons, nucleons, electrons/positrons) via the absorption (CC) 
or down-scattering ($E_\nu' < E_\nu$, NC).  
Note that the electrons, positrons and photons are 
regarded as a part of {\it matter}, 
being in thermal and chemical equilibrium with nucleons and 
nuclei in the supernova core, while the neutrinos are 
separately treated in the neutrino transfer calculations.  

\begin{figure}
\includegraphics[width=8cm]{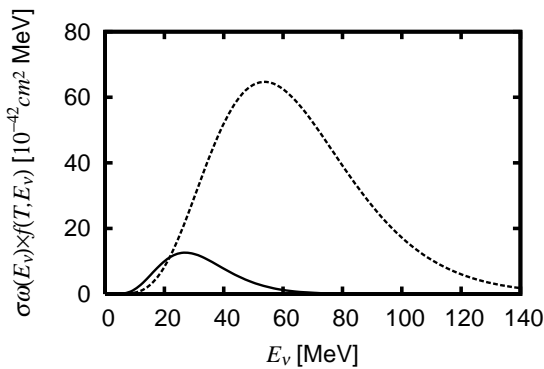}
\caption{Energy ($E_\nu$) dependence of the energy transfer cross section
for the CC $\nu d$ reaction multiplied by the Fermi-Dirac distribution
 function at $T_\nu=5$ MeV(solid curve) and $10$ MeV(dashed curve).}
\label{fig:edep}
\end{figure}

The integrand of the average energy transfer cross sections,
\begin{eqnarray}
 f(T_\nu,E_\nu) \sigma\omega(E_\nu) \ ,
\end{eqnarray}
for $\nu$ CC is plotted in Fig.~\ref{fig:edep} as a function of the
incident neutrino energy, $E_\nu$, at $T_\nu=5$ and $10$MeV.
The main contribution for the average cross section
is from $E_\nu \sim 20 (60)$MeV at $T_\nu=5 (10)$ MeV.
We note that 
the contribution from the high energy tail 
of $f\sigma\omega$ is appreciable.

\begin{figure}
\includegraphics[width=8cm]{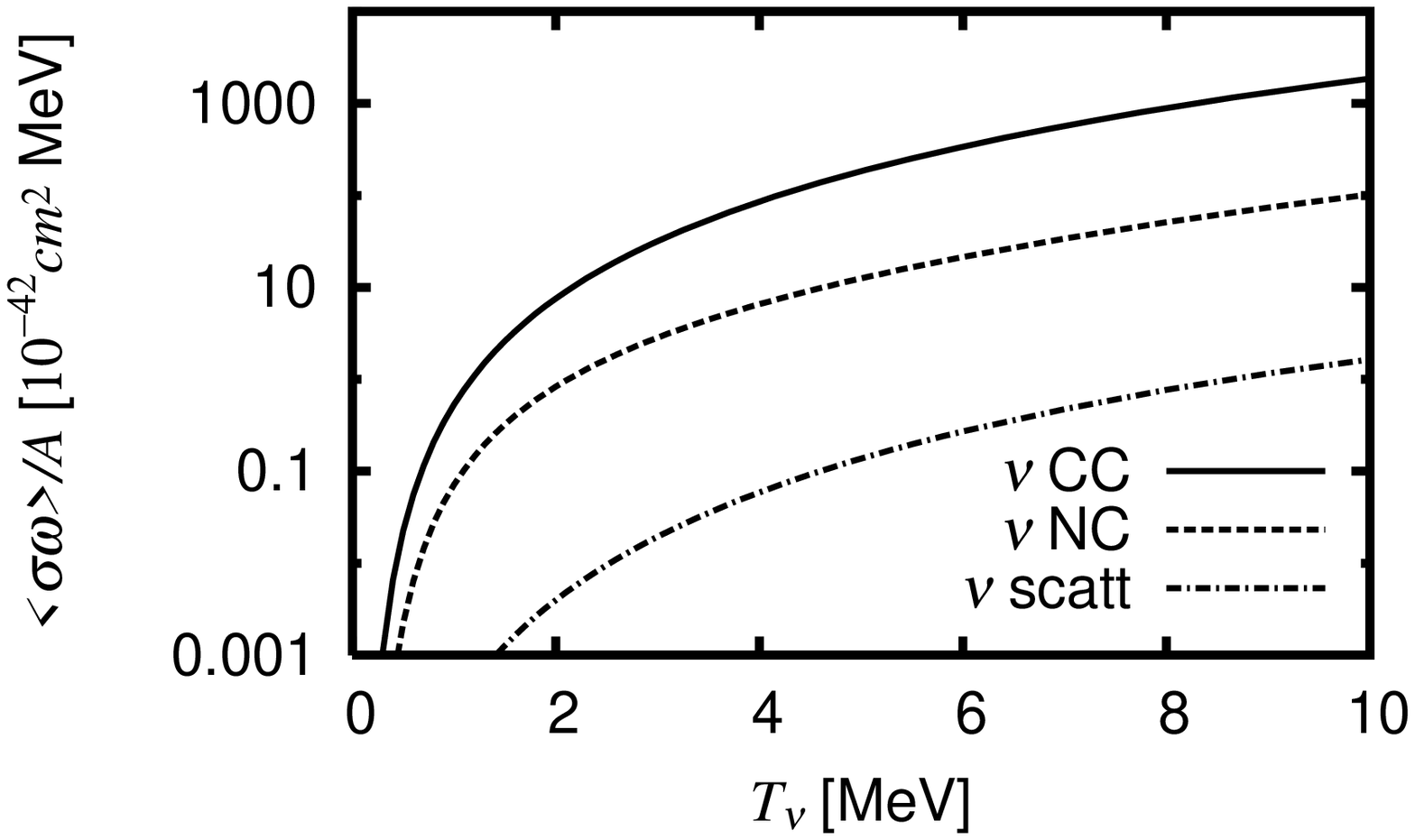}
\includegraphics[width=8cm]{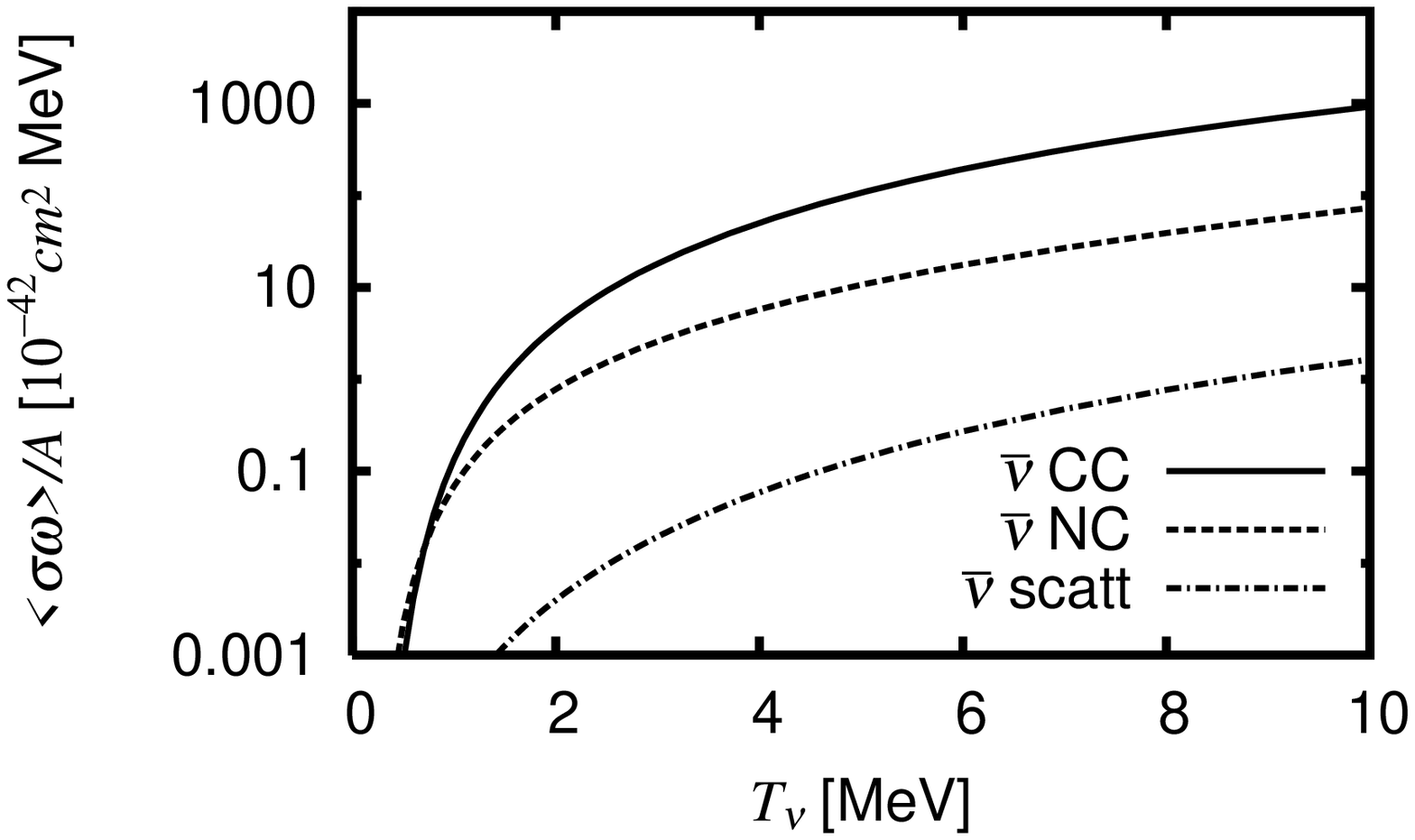}
\hspace*{2cm}(a)\hspace*{6cm}(b)
\caption{
Thermal average of energy transfer cross sections.
The solid and dotted curves in (a) ((b)) show the
cross sections for $\nu_e d \rightarrow e^- p p$($\nu$CC) and 
$\nu d \rightarrow \nu p n$($\nu$NC)
($\bar{\nu}_e d \rightarrow e^+ n n$($\bar{\nu}$CC) and
$\bar{\nu} d \rightarrow \bar{\nu} p n$($\bar{\nu}$NC)),
respectively. The dot-dashed curve in (a) and (b) shows
cross section for the elastic $\nu d$ scattering.
}
\label{fig:etrans}
\end{figure}

Finally our results on the thermal average of the energy 
transfer cross sections are shown in Fig. \ref{fig:etrans} and Table I.
The largest cross section is due to $\nu$CC which is almost the
same magnitude as $\bar{\nu}$CC. The CC cross sections
are an order of magnitude larger than the NC cross sections.
We remark that the $\nu d$ cross sections are appreciable 
even at relatively low temperature beyond the breakup threshold energy.  
This is a characteristic feature of the $\nu d$ reaction involving
the breakup of the weakly bound state with $2.2$MeV binding energy.
The rapid increase of the cross section at low energy makes the
deuteron more preferable target than the other nuclear species
in the supernova environment.

\begin{figure}
\includegraphics[width=8cm]{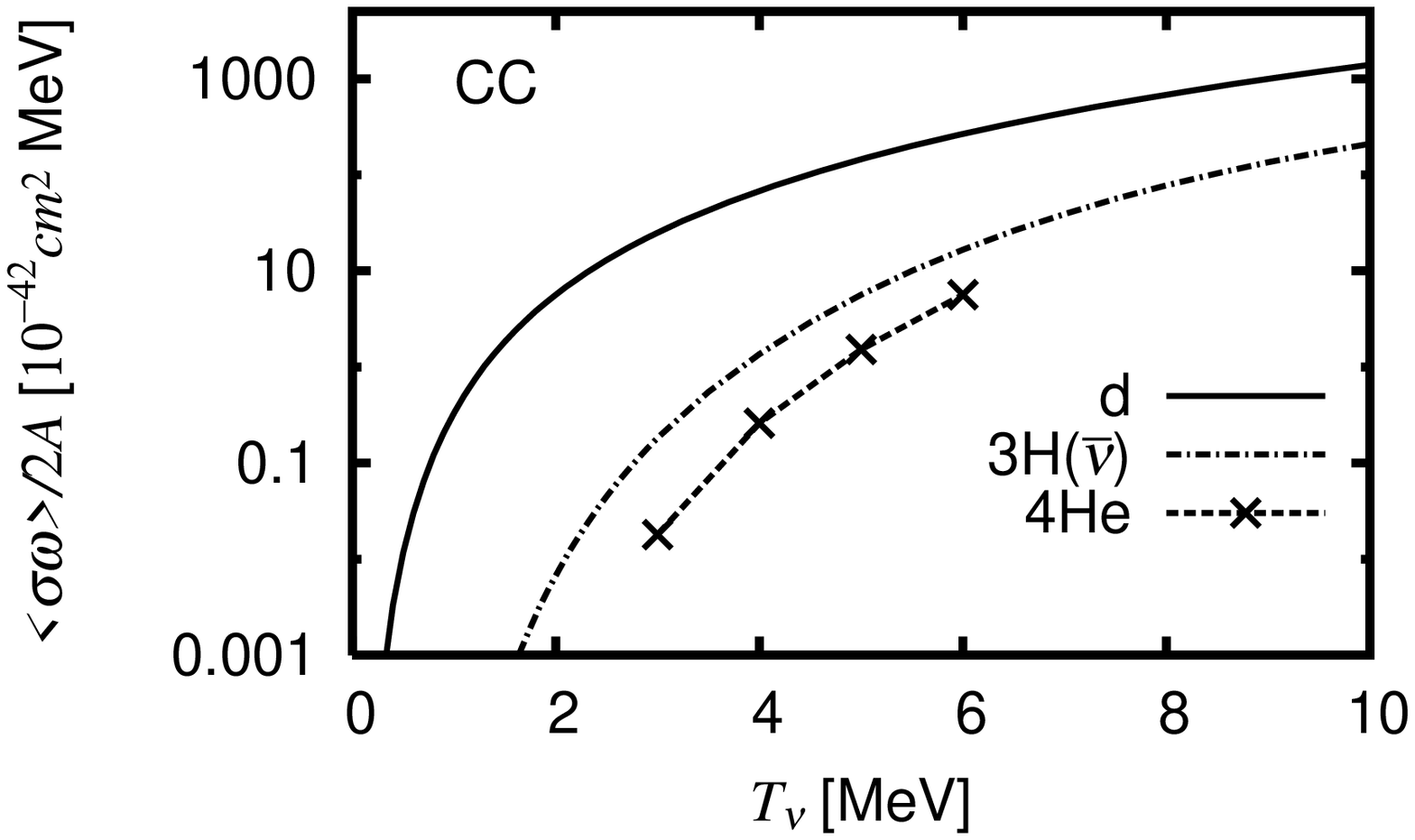}
\includegraphics[width=8cm]{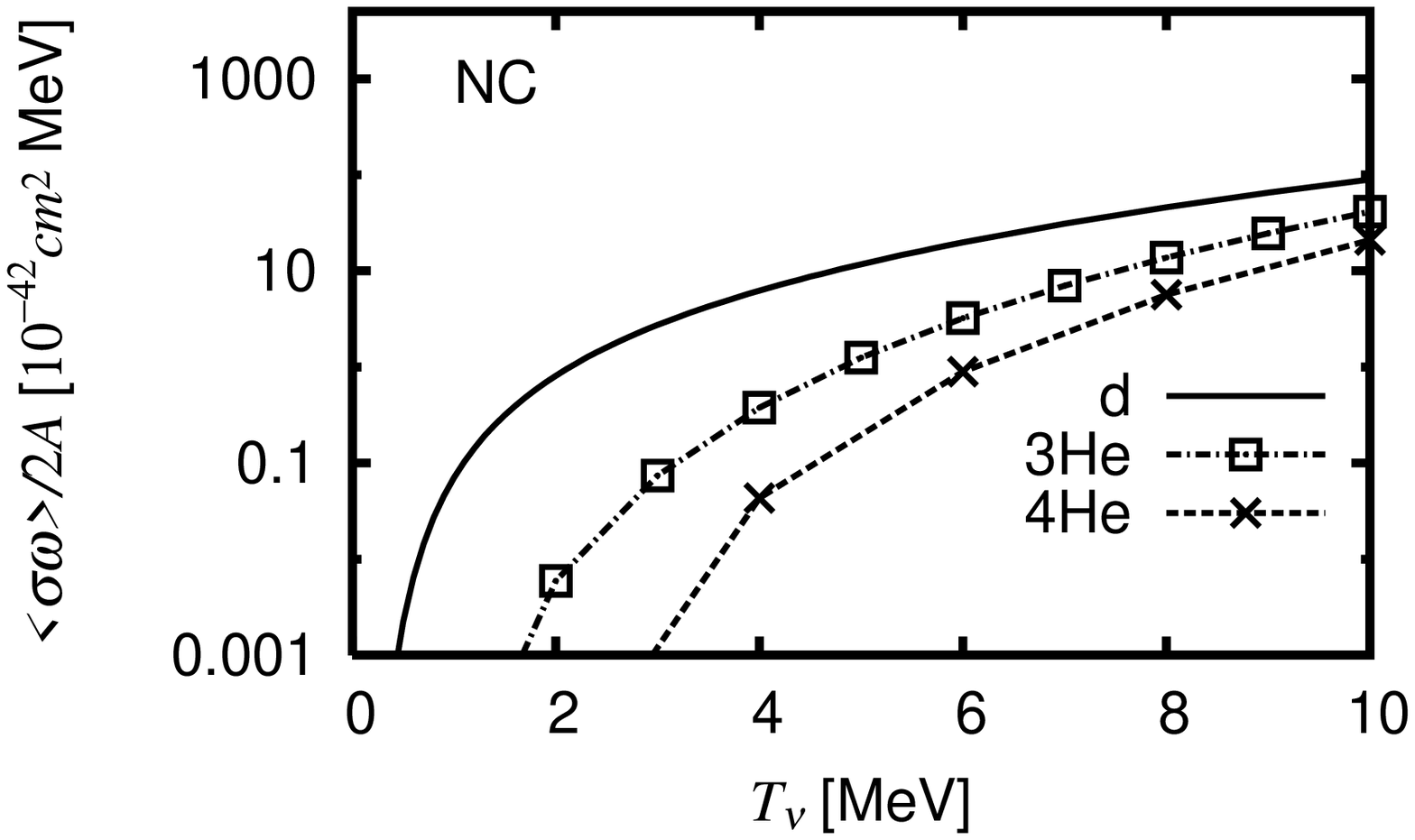}
\caption{Averaged energy transfer cross sections
in unit of $ 10^{-42}$ MeV cm$^2 $.
The solid, dashed and dash-dotted curves show
the $\nu d$, $\nu$-$^{4}$He and $\nu$-$^3$H (left panel) 
and $\nu$-$^{3}$He (right panel) cross sections, respectively.
(See the main text for the references on A=3, 4 nuclei cross sections.)
}
\label{fig:ave_etrans}
\end{figure}

In order to discuss the contribution of the deuteron
in the neutrino heating mechanism through a comparison 
with other neutrino-nucleus processes, 
we evaluate the average energy transfer rate per nucleon 
through the neutrino and anti-neutrino reactions defined by
\begin{eqnarray}
<\sigma \omega>^{CC} & = & \frac{1}{2A}
{} [<\sigma \omega>_{\nu CC} + <\sigma \omega>_{\bar{\nu} CC}]\ , \\
<\sigma \omega>^{NC} & = & \frac{1}{2A}
{} [<\sigma \omega>_{\nu NC} + <\sigma \omega>_{\bar{\nu} NC} 
+ <\sigma \omega>_{\nu scatt} +<\sigma \omega>_{\bar{\nu} scatt}]\ ,
\end{eqnarray}
where the factor 1/2 comes from the average over the neutrino and anti-neutrino.  
We show the calculated average energy transfer rate in Fig. \ref{fig:ave_etrans}.  
For comparison we show 
the average energy transfer rate for A=3, 4 nuclei.  
For the CC reactions, the $\bar{\nu}$-$^3$H rate 
is calculated from $\bar{\nu}$-$^3$H cross section given in Ref. \cite{arcones} 
and the $\nu$-$^{4}$He rate is taken from Ref. \cite{haxton}.  
For the NC reactions,
we take the values for the ${\nu}$-$^3$He rate from Ref. \cite{cghsb} and 
the ${\nu}$-$^4$He rate from Ref. \cite{gb}.
It is remarkable that
the average energy transfer cross sections for the $\nu d$ reaction 
is significantly larger than those for $^{3}$H, $^{3}$He and $^{4}$He 
for the temperature range relevant to the supernova core.  
The $\nu d$ rate is much larger at low temperature 
because of the low threshold energy of the deuteron as discussed previously.
In the case of the CC reaction
the $\nu d$ rate dominates, being larger than those for A=3, 4 nuclei 
by orders of magnitude.

The large energy transfer rate of the deuteron 
at the wide range of the temperature suggests that 
the deuteron is potentially an important target 
for the neutrino heating mechanism in supernovae.  
The average energy transfer cross section 
for the CC reaction is sufficiently large to yield a
significant heating rate for moderate neutrino temperatures.  
For example, the average energy transfer cross section 
for the CC reaction is larger than 10$^{-40}$ cm$^2$MeV
at $T_\nu \ge 4$ MeV, and gives a significant contribution 
to the heating rate in addition to the contribution of 
the neutrino-nucleon reactions.  
According to the expressions of the heating rate given by Haxton \cite{haxton}, 
the heating rate due to the $\nu d$ reactions 
for $\nu_e$ and $\bar{\nu}_e$ 
amounts to 99.1 MeV/sec per nucleon, assuming $T_{\nu_{e}}$=$T_{\bar{\nu}_{e}}$=5 MeV.  
The heating rate due to the $\nu d$  reactions for $\mu$ and
 $\tau$ type neutrino
is 55.6 MeV/sec per nucleon at $T_{\nu_{\mu}}$=10 MeV.  
These values are 25--44$\%$ of the heating rate, 223 MeV/sec per nucleon, 
due to the $\nu$-nucleon reactions at the same neutrino temperature of 5 MeV 
for $\nu_{e}$/$\bar{\nu_{e}}$.  
In the above estimate, we assumed values of the neutrino luminosities 
(L$_\nu$=10$^{52}$ erg/sec) and the distance from the center (100 km)
in the expression for the heating rate given in Ref.~\cite{haxton}.
As the neutrino temperature is higher so does the heating rate,
which makes the $\nu d$ reactions more important.  

We remark that the deuteron as well as other light elements 
can appear in hot and dense matter in the heating region 
between the proto-neutron star and the shock wave.  
The Fe-group nuclei falling from the outer layer 
are dissociated into $^{4}$He and then nucleons 
at the shock wave.  
Through this dissociation, light elements ($A \le 3$) 
including the deuteron can appear naturally 
in the hot environment \cite{cghsb,sr}.  
Although the nucleons, having larger cross sections 
than those for nuclei, are the major targets of the neutrino
after the complete dissociation, 
the deuteron and other light elements 
can contribute to the neutrino heating 
as an extra absorber of the neutrino.  

Light elements can be dominant targets 
in the case of enlarged shock radius 
due to the hydrodynamical SASI instability 
as shown by Ohnishi et al. \cite{ohnishi} 
in multi-dimensional supernova simulations.  
In such case, the range of density and temperature 
is favorable to have  abundant $^{4}$He with
enough advection time to have neutrino absorptions.
In their study, the $\nu$-$^{4}$He reactions 
may be crucial for the revival of shock wave 
in the marginal condition, but they require  
the enhanced $\nu$-$^{4}$He reaction rates.  
Since the $\nu d$ reaction rates is much 
larger than those of $^{4}$He as we have seen, 
the deuteron may play a similar role instead of $^{4}$He
during its dissociation.  
It would be interesting to explore the effect 
of $\nu d$ reactions in the multi-dimensional 
supernova simulations 
with a detailed evaluation of mixture of light 
elements.  
The environment with an elongated shock wave 
due to the SASI instability may prefer
to have abundant deuterons together with $^{4}$He.  
It has been shown that the deuteron appears ($\sim1\%$)
in the heating region behind the stalled shock wave 
in a realistic snapshot of central core after the bounce 
in the spherical supernova simulations \cite{sr}.  

\begin{table}
\begin{tabular}{c|ccccc} \hline \hline 
$T_{\nu}$(MeV) 
   & $\nu$ CC          & $\bar{\nu}$CC       & $\nu$ NC              & $\bar{\nu}$ NC      & $\nu$ Scatt. \\ \hline
 1 & $1.06\times 10^0$ & $2.68\times 10^{-1}$& $1.43\times 10^{-1}$  &  $1.38\times 10^{-1}$ &   $5.11\times 10^{-4}$\\
 2 & $1.51\times 10^1$ & $7.50\times 10^0$   & $1.66\times 10^0$     &  $1.56\times 10^0$    &   $7.95\times 10^{-3}$\\
 3 & $6.37\times 10^1$ & $3.63\times 10^1$   & $5.68\times 10^0$     &  $5.17\times 10^0$    &   $3.89\times 10^{-2}$\\
 4 & $1.71\times 10^2$ & $1.01\times 10^2$   & $1.31\times 10^1$     &  $1.15\times 10^1$    &   $1.18\times 10^{-1}$\\
 5 & $3.67\times 10^2$ & $2.14\times 10^2$   & $2.50\times 10^1$     &  $2.13\times 10^1$    &   $2.75\times 10^{-1}$\\
 6 & $6.79\times 10^2$ & $3.87\times 10^2$   & $4.28\times 10^1$     &  $3.51\times 10^1$    &   $5.42\times 10^{-1}$\\
 7 & $1.14\times 10^3$ & $6.29\times 10^2$   & $6.80\times 10^1$     &  $5.37\times 10^1$    &   $9.51\times 10^{-1}$\\
 8 & $1.78\times 10^3$ & $9.49\times 10^2$   & $1.02\times 10^2$     &  $7.80\times 10^1$    &   $1.53\times 10^0$\\
 9 & $2.64\times 10^3$ & $1.35\times 10^3$   & $1.48\times 10^2$     &  $1.09\times 10^2$    &   $2.31\times 10^0$\\
10 & $3.73\times 10^3$ & $1.85\times 10^3$   & $2.06\times 10^2$     &  $1.46\times 10^2$    &   $3.30\times 10^0$\\ 
\hline\hline
\end{tabular}
\caption{Averaged energy transfer cross sections
 $<\sigma\omega>$ 
in unit of 10$^{-42}$ MeV cm$^2$. }
\end{table}

Further studies on the suitable condition to have a sufficient
amount of targets and a long enough reaction time periods
are necessary to firm the possible role of neutrino-deuteron 
interactions for the success of supernova explosions.
The systematic study of the abundance of light elements 
in hot and dense matter is now under way \cite{sr2}. 

In summary:
the neutrino-deuteron reactions may have a potential impact 
on the neutrino heating mechanism in core-collapse supernovae.  
The cross section is large even at low neutrino energies 
relevant to supernovae 
because of the low breakup threshold for the bound state.  
The energy transfer rate through the $\nu d$ reactions 
is much larger than those of $^{3}$H, $^{3}$He and $^{4}$He 
even at low neutrino temperature.  
If deuterons appear abundantly in the heating region, 
as $^{4}$He in the SASI instability, 
they can contribute to the extra heating, 
which assists the revival of stalled shock wave.  
It is interesting to explore whether the $\nu d$
heating can affect the dynamics of shock wave 
in realistic supernova simulations which take into account 
the mixture of light elements and the associated neutrino 
reactions.  

The evaluated data of the $\nu d$ energy transfer rate 
will be available on the web site for the neutrino deuteron reactions\cite{web}.  
The data can be used to implement these processes 
into the heating rate in supernova simulations.  
Detailed neutrino differential cross sections 
based on the current theoretical approach
can be provided for numerical simulations of the neutrino transfer 
which require the angle and energy variations.
It will be also interesting to study the $\nu d$
processes around the surface of proto-neutron stars 
and its influence on the neutrino nucleosynthesis.  
The other channels of weak processes on the deuteron
are now under investigation.  

\begin{acknowledgments}

This work is supported by the Japan Society
 for the Promotion of Science (JSPS) 
Grant-in-Aid for Scientific Research (20540270) and 
is partially supported by the JSPS Grants-in-Aid for Scientific Research 
(18540291, 18540295, 19540252) 
and the Grant-in-Aid for Scientific Research on Innovative Areas 
of the MEXT (20105004).
K. S. is grateful to N. Ohnishi, S. Yamada, H. Suzuki and G. R{\"o}pke 
for the collaborations on the supernova simulations and the light elements 
in dense matter.  
We would like to thank D. Gazit for providing us the numerical data 
of energy transfer rates.  

\end{acknowledgments}

\end{document}